\documentclass[12pt]{article}

\usepackage{epsf}
\usepackage{rotating}
\usepackage{color}
\newcommand{\beq}{\begin{equation}}
\newcommand{\eeq}{\end{equation}}
\newcommand{\bea}{\begin{eqnarray}}
\newcommand{\eea}{\end{eqnarray}}
\newcommand{\bed}{\begin{displaymath}}
\newcommand{\eed}{\end{displaymath}}

\catcode`\@=11

\def\@citex[#1]#2{\if@filesw\immediate\write\@auxout{\string\citation{#2}}\fi
  \def\@citea{}\@cite{\@for\@citeb:=#2\do
    {\@citea\def\@citea{,\penalty\@m}\@ifundefined
       {b@\@citeb}{{\bf ?}\@warning
       {Citation `\@citeb' on page \thepage \space undefined}}%
\hbox{\csname b@\@citeb\endcsname}}}{#1}}

\def\citer{\@ifnextchar [{\@tempswatrue\@citexr}{\@tempswafalse\@citexr[]}}
\def\@citexr[#1]#2{\if@filesw\immediate\write\@auxout{\string\citation{#2}}\fi
  \def\@citea{}\@cite{\@for\@citeb:=#2\do
    {\@citea\def\@citea{--\penalty\@m}\@ifundefined
       {b@\@citeb}{{\bf ?}\@warning
       {Citation `\@citeb' on page \thepage \space undefined}}%
\hbox{\csname b@\@citeb\endcsname}}}{#1}}
\catcode`\@=12

\begin{document}

\renewcommand{\thefootnote}{\fnsymbol{footnote}}

\renewcommand{\thefootnote}{\arabic{footnote}}

\setcounter{footnote}{0}

\vspace*{-2cm}

\begin{flushright}
CERN--TH--2020--139 \\
IFIC/20--42 \\
FTUV--20--0823 \\
KA--TP--11--2020 \\
PSI--PR--20--13
\end{flushright}
\begin{center}
{\large \sc $gg\to HH$: Combined Uncertainties}
\\[0.5cm]

J.~Baglio$^1$, F.~Campanario$^{2,3}$, S.~Glaus$^{3,4}$,
M.~M\"uhlleitner$^3$, J.~Ronca$^2$ and M.~Spira$^5$ \\
\end{center}

\noindent
{\it $^1$ Theory Physics Department, CERN, CH--1211 Geneva 23,
Switzerland} \\[0.1cm]
{\it $^2$ Theory Division, IFIC, University of Valencia-CSIC, E--46980
Paterna, Valencia, Spain} \\[0.1cm]
{\it $^3$ Institute for Theoretical Physics, Karlsruhe Institute of
Technology, D--76131 Karlsruhe, Germany}\\[0.1cm]
{\it $^4$ Institute for Nuclear Physics, Karlsruhe Institute of
Technology, D--76344 Karlsruhe, Germany} \\[0.1cm]
{\it $^5$ Paul Scherrer Institut, CH--5232 Villigen PSI, Switzerland}

\begin{abstract}
\noindent
In this note we discuss the combination of the usual renormalization and
factorization scale uncertainties of Higgs-pair production via gluon
fusion with the novel uncertainties originating from the scheme and
scale choice of the virtual top mass. Moreover, we address the
uncertainties related to the top-mass definition for different values of
the trilinear Higgs coupling and their combination with the other
uncertainties.
\end{abstract}

\section{Introduction}
Higgs-boson pair production will allow for the first time to probe the
trilinear Higgs self-coupling directly and thus to determine the first
part of the Higgs potential as the origin of electroweak symmetry
breaking. The dominant Higgs pair production mode is gluon fusion $gg\to
HH$ that is loop-induced at leading order (LO), mediated by top and to a
much lesser extent bottom loops \cite{gg2hhlo}. The total gluon-fusion
cross section is about three orders of magnitude smaller than the
corresponding single-Higgs production cross section \cite{hhrev}. The
dependence of the gluon-fusion cross section on the trilinear Higgs
self-coupling $\lambda$ around the Standard-Model (SM) value is
approximately given by $\Delta\sigma/\sigma \sim -\Delta\lambda/\lambda$
so that the uncertainties of the cross section are immediately
translated into the uncertainty of the extracted trilinear
self-coupling. In order to reduce the uncertainties of the cross section
higher-order corrections are required. The next-to-leading-order (NLO)
QCD corrections have first been obtained in the heavy-top limit (HTL)
\cite{gg2hhnlo} supplemented by a large top-mass expansion
\cite{gg2hhnloexp} and the inclusion of the full real corrections
\cite{gg2hhnloreal}. Meanwhile, the full NLO calculation including the
full top-mass dependence has become available
\cite{gg2hha,gg2hhb,gg2hhc} showing a 15\%-difference to the result
obtained in the HTL for the total cross section. For the distributions
the differences can reach 20--30\% for large invariant Higgs pair
masses. The full NLO results have been confirmed by suitable expansion
methods \cite{expansion}. Within the HTL the next-to-NLO (NNLO)
\cite{gg2hhnnlo} and next-to-NNLO (N$^3$LO) \cite{gg2hhn3lo} QCD
corrections have been derived and raise the cross section by a moderate
amount of 20--30\% in total. The complete QCD corrections increase the
cross section by more than a factor of two. Quite recently, the full NLO
result and the NNLO corrections in the HTL have been combined in a fully
exclusive Monte Carlo program \cite{gg2hhmc} (including the mass effects
of the one-loop double-real contributions at NNLO) that is publicly
available\footnote{The approach of Ref.~\cite{gg2hhmc} is called
NNLO$_{FTapprox}$.}. Moreover, the matching of the full NLO results to
parton showers has been performed \cite{gg2hhps} so that there are
complete NLO event generators.

\section{Uncertainties}
The usual renormalization and factorization scale uncertainties at NLO
amount to about 10--15\% \cite{gg2hha,gg2hhc},
\begin{eqnarray}
\sqrt{s} = 13~{\rm TeV}: \quad
\sigma_{tot} & = & 27.73(7)^{+13.8\%}_{-12.8\%}~{\rm fb}, \nonumber \\
\sqrt{s} = 14~{\rm TeV}: \quad
\sigma_{tot} & = & 32.81(7)^{+13.5\%}_{-12.5\%}~{\rm fb}, \nonumber \\
\sqrt{s} = 27~{\rm TeV}: \quad
\sigma_{tot} & = & 127.0(2)^{+11.7\%}_{-10.7\%}~{\rm fb}, \nonumber \\
\sqrt{s} = 100~{\rm TeV}: \quad
\sigma_{tot} & = & 1140(2)^{+10.7\%}_{-10.0\%}~{\rm fb},
\label{eq:signlo}
\end{eqnarray}
where $s$ denotes the squared center-of-mass energy and $\sigma_{tot}$
the total cross section.  The numbers in brackets are the numerical
integration errors and the upper and lower per-centage entries denote
the combined renormalization and factorization scale uncertainties. They
have been obtained by a (7-point) variation of the renormalization and
factorization scales $\mu_R, \mu_F$ by a factor of two around the
central (dynamical) scale $\mu_0=M_{HH}/2$, where $M_{HH}$ denotes the
invariant Higgs-pair mass. The numbers of Eq.~(\ref{eq:signlo}) have
been obtained for a top pole mass of $m_t=172.5$ GeV, a Higgs mass of
$M_H=125$ GeV and PDF4LHC parton distribution functions (PDFs)
\cite{pdf4lhc}. However, in addition to the scale dependence of the
strong coupling constant and PDFs, the virtual
top mass is subject to a scheme and scale dependence, too. This involves
the top mass included in the top Yukawa coupling as well as the top mass
entering the virtual top propagators.

The (central) numbers of Eq.~(\ref{eq:signlo}) are obtained in terms of
the top pole mass. In order to derive the corresponding results with the
top $\overline{\rm MS}$ mass $\overline{m}_t$ for both the Yukawa
coupling and propagator mass we use the N$^3$LO relation between the
pole and $\overline{\rm MS}$ mass
\begin{eqnarray}
{\overline{m}}_{t}(m_{t}) & = & \frac{m_{t}}{\displaystyle 1+\frac{4}{3}
\frac{\alpha_{s}(m_t)}{\pi} + K_2
\left(\frac{\alpha_s(m_t)}{\pi}\right)^2 + K_3
\left(\frac{\alpha_s(m_t)}{\pi}\right)^3}
\label{eq:mspole}
\end{eqnarray}
with $K_2\approx 10.9$ and $K_3 \approx 107.11$. The scale dependence of
the $\overline{\rm MS}$ mass is treated at
next-to-next-to-next-to-leading logarithmic level (N$^3$LL),
\begin{eqnarray}
{\overline{m}}_{t}\,(\mu_t)&=&{\overline{m}}_{t}\,(m_{t})
\,\frac{c\,[\alpha_{s}\,(\mu_t)/\pi ]}{c\, [\alpha_{s}\,(m_{t})/\pi ]}
\label{eq:msbarev}
\end{eqnarray}
with the coefficient function \cite{anomass}
\begin{eqnarray}
c(x)=\left(\frac{7}{2}\,x\right)^{\frac{4}{7}} \, [1+1.398x+1.793\,x^{2}
- 0.6834\, x^3]\, .
\end{eqnarray}
This introduces a new scale $\mu_t$, the dependence on which induces an
additional uncertainty. For large values of the invariant Higgs-pair
mass, the high-energy expansion of the virtual form factors clearly
favors the dynamical scale choice $\mu_t \sim M_{HH}$
\cite{gg2hhc,gg2hhapprox}.

The scale dependence of the total and differential Higgs-pair production
cross section on $\mu_t$ drops by roughly a factor of two from LO to NLO
as explicitly described in Ref.~\cite{gg2hhc}. The procedure to obtain
the associated uncertainties is to take the envelope of the different
predictions with the top pole mass and the $\overline{\rm MS}$ mass
$\overline{m}_t (\mu_t)$ at the scale $\mu_t = \overline{m}_t$ and
varying it between $M_{HH}/4$ and $M_{HH}$ (i.e.~a factor of 2 around
the central renormalization and factorization scale
$\mu_R=\mu_F=M_{HH}/2$) for each $M_{HH}$ bin and integrating the
maxima/minima eventually. At NLO we are left with the residual
uncertainties related to the top-mass scheme and scale choice
\cite{gg2hhb,gg2hhc},
\begin{eqnarray} 
\sqrt{s} = 13~{\rm TeV}: \quad 
\sigma_{tot} & = & 27.73(7)^{+4\%}_{-18\%}~{\rm fb}, \nonumber \\
\sqrt{s} = 14~{\rm TeV}: \quad
\sigma_{tot} & = & 32.81(7)^{+4\%}_{-18\%}~{\rm fb}, \nonumber \\
\sqrt{s} = 27~{\rm TeV}: \quad
\sigma_{tot} & = & 127.8(2)^{+4\%}_{-18\%}~{\rm fb}, \nonumber \\
\sqrt{s} = 100~{\rm TeV}: \quad
\sigma_{tot} & = & 1140(2)^{+3\%}_{-18\%}~{\rm fb}
\label{eq:mtunc}
\end{eqnarray}
A further reduction of these uncertainties can only be achieved by the
determination of the full mass effects at NNLO which is beyond the
state of the art\footnote{Due to the moderate size of the NNLO
corrections a reduction of these uncertainties by a factor $\sim$ 3--4
may be expected by the NNLO mass effects.}. Since these uncertainties
are sizeable, the question arises of how to combine them with the other
renormalization and factorization scale uncertainties of
Eq.~(\ref{eq:signlo}).

The interplay of the different uncertainties of
Eqs.~(\ref{eq:signlo},\ref{eq:mtunc}) at NLO is very simple,
i.e.~defining the envelope of all uncertainties leads to a {\it linear}
addition of the renormalization and factorization scale uncertainties of
Eq.~(\ref{eq:signlo}) and the top-mass scheme and scale uncertainties of
Eq.~(\ref{eq:mtunc}), since the latter turn out to be (nearly)
independent of the renormalization and factorization scale choices. This
statement has been evaluated up to NLO explicitly.

The presently recommended predictions and uncertainties are based on the
work of Ref.~\cite{gg2hhmc}. This work includes the NNLO QCD corrections
in the HTL combined with the full mass effects of the
LO and NLO predictions. Moreover, the work includes the full mass
dependence of the one-loop double-real corrections at NNLO. The central
values and residual renormalization and factorization scale
uncertainties of this approach are given by \cite{gg2hhmc,whitepaper}
\begin{eqnarray} 
\sqrt{s} = 13~{\rm TeV}: \quad 
\sigma_{tot} & = & 31.05^{+2.2\%}_{-5.0\%}~{\rm fb}, \nonumber \\
\sqrt{s} = 14~{\rm TeV}: \quad
\sigma_{tot} & = & 36.69^{+2.1\%}_{-4.9\%}~{\rm fb}, \nonumber \\
\sqrt{s} = 27~{\rm TeV}: \quad
\sigma_{tot} & = & 139.9^{+1.3\%}_{-3.9\%}~{\rm fb}, \nonumber \\
\sqrt{s} = 100~{\rm TeV}: \quad
\sigma_{tot} & = & 1224^{+0.9\%}_{-3.2\%}~{\rm fb}\, .
\label{eq:uncstatus}
\end{eqnarray}
These uncertainties will be further reduced by consistently including the
novel N$^3$LO corrections in the HTL \cite{gg2hhn3lo}.

\section{Combination of Uncertainties}
In order to find a proper scheme to combine the renormalization and
factorization scale uncertainties of Eq.~(\ref{eq:uncstatus}) and the
uncertainties originating from the top-mass scheme and scale choice of
Eq.~(\ref{eq:mtunc}) we have to consider the systematics of these
uncertainties in more detail. Each perturbative order of the total (and
differential) cross section in QCD can be decomposed in two different
pieces of the corrections,
\begin{eqnarray}
d \sigma_n & = & \sum_{i=0}^n d\sigma^{(i)} \nonumber \\
d \sigma_n & = & d \sigma_{n-1} \times (K^{(n)}_{SVC} + K^{(n)}_{rem})
\end{eqnarray}
where $d \sigma_n$ denotes the $n$'th-order-corrected differential cross
section, $d\sigma^{(i)}$ the $i$'th-order correction, $K^{(n)}_{SVC}$
the universal part of the soft+virtual+collinear corrections and
$K^{(n)}_{rem}$ the remainder of the $n$'th-order corrections relative
to the previous order of the cross section. The (top-mass independent)
part $K^{(i)}_{SVC}$ is dominant for the first few orders, while the
moderate (top-mass dependent) remainder $K^{(i)}_{rem}$ only adds
10--15\% to the bulk of the corrections of $\sim 100\%$. The
soft+virtual corrections $K^{(i)}_{SVC}$ are basically the same for the
(subleading) mass-effects at all orders, too. Since these pieces are
part of the HTL at all perturbative orders the Born-improved
\cite{gg2hhnlo} and FTapprox \cite{gg2hhnloreal} approaches provide a
reasonable approximation of the total cross section within 10--15\% at
NLO. The mass effects at a given order are thus multiplied by the same
universal correction factors, too. In the same way, the uncertainties
originating from the mass effects are scaling with this dominant part of
the QCD corrections. This statement is explicitly corroborated by the
fact that the (Born-improved) HTL approximates the NLO cross section
within about 15\%, while the QCD corrections modify the cross section by
close to 100\%. Hence, at the state of the art, i.e.~full NLO and
NNLO\footnote{In the future, the novel N$^3$LO results will eventually
become part of the recommended values.} within the HTL with massive
refinements, the best procedure to combine the {\it relative}
uncertainties of Eqs.~(\ref{eq:mtunc}) and Eq.~(\ref{eq:uncstatus}) is
{\it linearly}.  This will be not only the most conservative approach,
but close to the final numbers in a sophisticated combined calculation
of the NNLO results in the HTL with the full NLO mass effects, i.e.~with
a negligible mismatch of the envelope from the linear
combination\footnote{Our approach is not meant to estimate the
uncertainties at full NNLO but the uncertainties at approximate NNLO
without the knowledge of the complete $m_t$-effects at NNLO.}.

\begin{figure}[hbtp]
\begin{center}

  \includegraphics[width=0.49\textwidth]{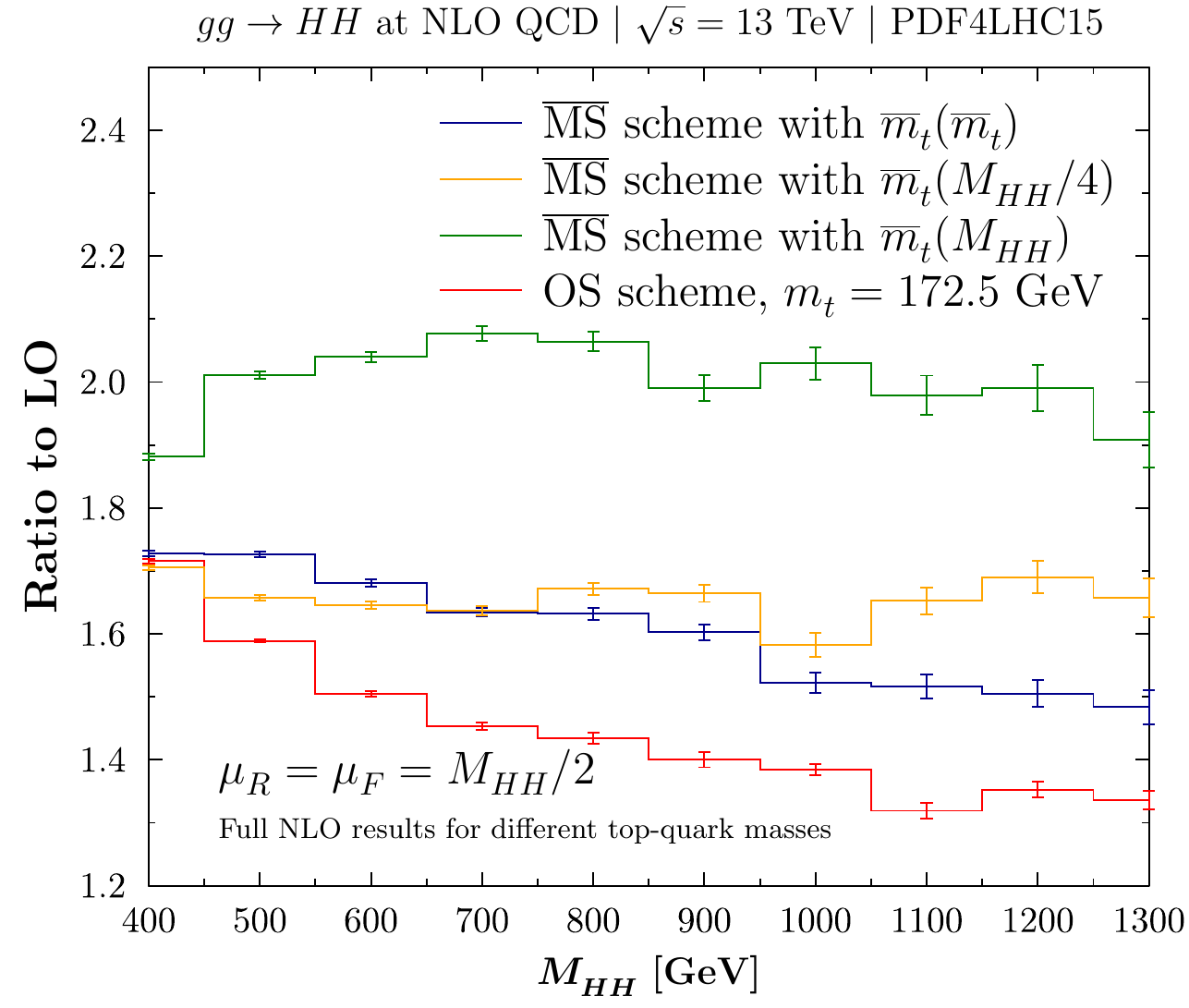}
  \includegraphics[width=0.49\textwidth]{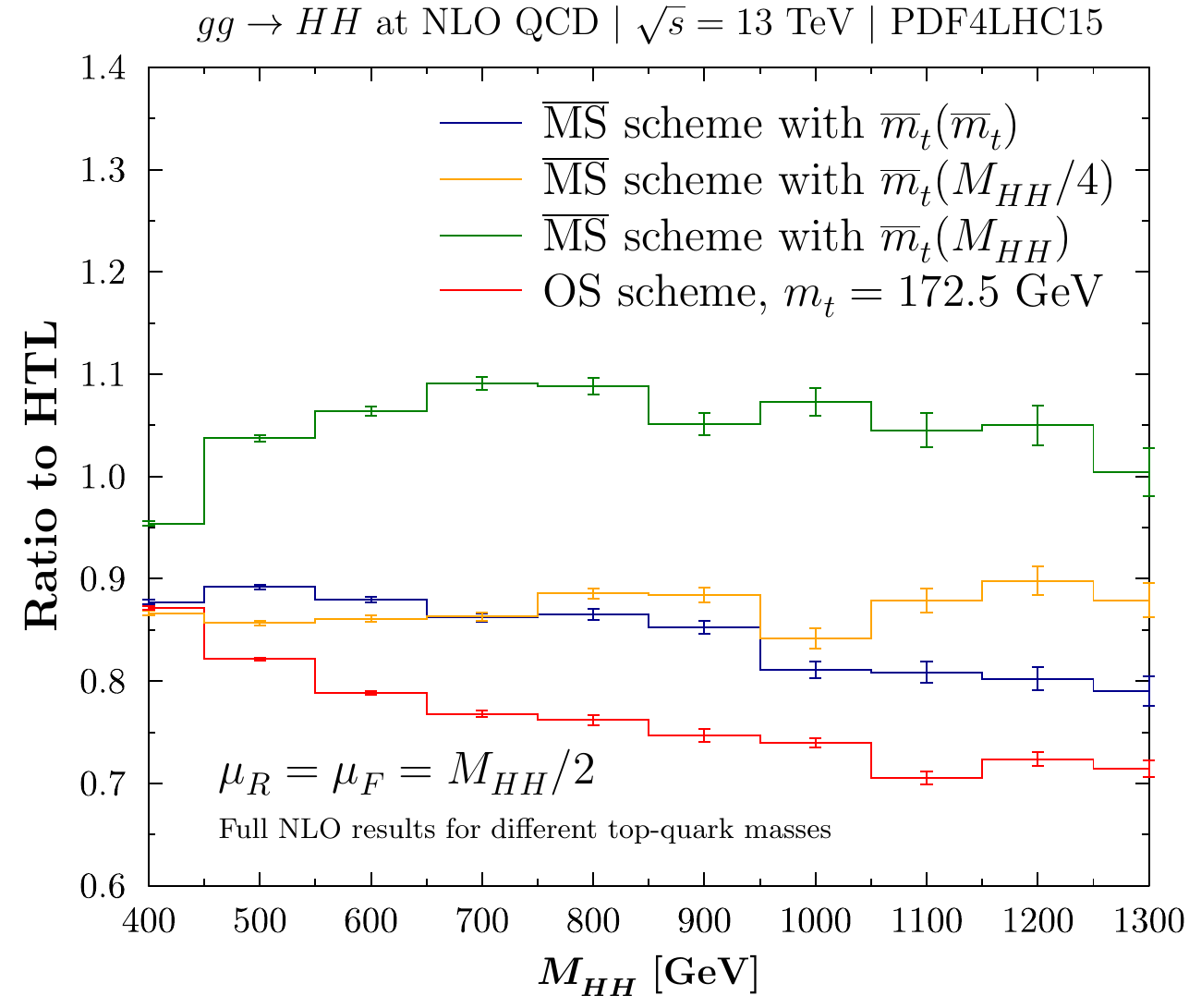}
%
\caption{\it Ratio of the full NLO QCD corrected differential cross
section to the LO one (left) and to the (Born-improved) NLO HTL (right)
for various definitions of the virtual top mass as a function of the
invariant Higgs-pair mass $M_{HH}$ for a c.m.~energy $\sqrt{s}=14$ TeV
and using PDF4LHC parton densities.}
\label{fg:ratio}
\end{center}
\end{figure}

This procedure results in the following combined uncertainties of
Eqs.~(\ref{eq:mtunc},\ref{eq:uncstatus}),
\begin{eqnarray} 
\sqrt{s} = 13~{\rm TeV}: \quad 
\sigma_{tot} & = & 31.05^{+6\%}_{-23\%}~{\rm fb}, \nonumber \\
\sqrt{s} = 14~{\rm TeV}: \quad
\sigma_{tot} & = & 36.69^{+6\%}_{-23\%}~{\rm fb}, \nonumber \\
\sqrt{s} = 27~{\rm TeV}: \quad
\sigma_{tot} & = & 139.9^{+5\%}_{-22\%}~{\rm fb}, \nonumber \\
\sqrt{s} = 100~{\rm TeV}: \quad
\sigma_{tot} & = & 1224^{+4\%}_{-21\%}~{\rm fb}
\end{eqnarray}
The central values of these numbers have been obtained by using the top
pole mass. In light of the findings of Refs.~\cite{gg2hhc,gg2hhapprox}
the preferred scale choice is $\mu_t\sim M_{HH}$ at large values of
$M_{HH}$ so that the choice of the top pole mass for the central
prediction can be questioned. However, for small values of $M_{HH}$
close to the production threshold the process is quite close to the HTL,
where the scale choice $\mu_t\sim m_t$ is the preferred one, since the
top mass constitutes the related matching scale. The scale choice
$\mu_t=m_t$ is implicitly involved in the top pole mass, too. A further
refinement of the proper scale choice for the virtual top mass would
require an interpolation between the different kinematical regimes that
would introduce a new uncertainty by itself. Such investigations are
beyond the scope of this note and all analyses so far. It should,
however, be noted that the relative NLO top-mass effects turn out to be
quite independent of $M_{HH}$ if the top mass is defined as the
$\overline{\rm MS}$ mass $\overline{m}_t(M_{HH}/4)$ as can be inferred
from Fig.~\ref{fg:ratio}, where we display the ratio of the NLO cross
section to the LO cross section\footnote{It should be noted that the
ratio to the LO cross section is not the consistently defined K factor.
The latter requires the LO cross section to be evaluated with LO
$\alpha_s$ and PDFs, while we use NLO quantities at LO, too, to show the
pure effects of the matrix elements.} and to the Born-improved HTL at
NLO (with the LO cross section determined in terms of the used top mass
definition) for various choices of the top mass. Adopting
$\overline{m}_t(M_{HH}/4)$ for the top mass the NLO mass effects range
between 10\% and 15\% for the whole range in $M_{HH}$ with a mild
dependence on the invariant Higgs-pair mass as can be inferred from the
ratio to the HTL. The ratio to the LO cross section develops a very flat
behaviour for this scale choice, too.

\section{Uncertainties for different Higgs self-interactions}
A variation of the trilinear Higgs coupling $\lambda$ modifies the
interplay between the LO box and triangle contributions that interfere
destructively for the SM case. One of the basic questions is what will
happen to the uncertainties for different values of $\lambda$. This can
be traced back to the approximately aligned uncertainties of the triangle
and box diagrams \cite{gg2hhc,houches}. The renormalization and
factorization scale uncertainties change by up to about 6\% at NLO for
large and small values of $\lambda$ \cite{whitepaper} such that the
change with respect to the central uncertainties of the SM value of
$\sim$ 10--15\% is of moderate size. In a similar way the uncertainties
originating from the scheme and scale choice of the top mass depend only
mildly on the trilinear coupling $\lambda$. Eq.~(\ref{eq:siglam_rf})
shows the central NNLO$_{FTapprox}$ predictions for the total cross
section for various choices of $\kappa_\lambda = \lambda/\lambda_{SM}$
for $\sqrt{s}=13$ TeV. The per-cent uncertainties display the usual
factorization and renormalization scale uncertainties \cite{hxswghh}.
\begin{eqnarray} 
\kappa_\lambda = -10: \quad 
\sigma_{tot} & = & 1680^{+3.0\%}_{-7.7\%}~{\rm fb}, \nonumber \\
\kappa_\lambda = -5: \quad 
\sigma_{tot} & = & 598.9^{+2.7\%}_{-7.5\%}~{\rm fb}, \nonumber \\
\kappa_\lambda = -1: \quad 
\sigma_{tot} & = & 131.9^{+2.5\%}_{-6.7\%}~{\rm fb}, \nonumber \\
\kappa_\lambda = 0: \quad 
\sigma_{tot} & = & 70.38^{+2.4\%}_{-6.1\%}~{\rm fb}, \nonumber \\
\kappa_\lambda = 1: \quad 
\sigma_{tot} & = & 31.05^{+2.2\%}_{-5.0\%}~{\rm fb}, \nonumber \\
\kappa_\lambda = 2: \quad 
\sigma_{tot} & = & 13.81^{+2.1\%}_{-4.9\%}~{\rm fb}, \nonumber \\
\kappa_\lambda = 2.4: \quad 
\sigma_{tot} & = & 13.10^{+2.3\%}_{-5.1\%}~{\rm fb}, \nonumber \\
\kappa_\lambda = 3: \quad 
\sigma_{tot} & = & 18.67^{+2.7\%}_{-7.3\%}~{\rm fb}, \nonumber \\
\kappa_\lambda = 5: \quad 
\sigma_{tot} & = & 94.82^{+4.9\%}_{-8.8\%}~{\rm fb}, \nonumber \\
\kappa_\lambda = 10: \quad 
\sigma_{tot} & = & 672.2^{+4.2\%}_{-8.5\%}~{\rm fb}
\label{eq:siglam_rf}
\end{eqnarray}
These predictions for the cross sections have been obtained by adopting
the top pole mass for the LO and higher-order contributions. Modifying
the scheme and scale choice of the top mass according to the SM analysis
we end up with the additional uncertainties at NLO
\begin{eqnarray} 
\kappa_\lambda = -10: \quad 
\sigma_{tot} & = & 1438(1)^{+10\%}_{-6\%}~{\rm fb}, \nonumber \\
\kappa_\lambda = -5: \quad 
\sigma_{tot} & = & 512.8(3)^{+10\%}_{-7\%}~{\rm fb}, \nonumber \\
\kappa_\lambda = -1: \quad 
\sigma_{tot} & = & 113.66(7)^{+8\%}_{-9\%}~{\rm fb}, \nonumber \\
\kappa_\lambda = 0: \quad 
\sigma_{tot} & = & 61.22(6)^{+6\%}_{-12\%}~{\rm fb}, \nonumber \\
\kappa_\lambda = 1: \quad 
\sigma_{tot} & = & 27.73(7)^{+4\%}_{-18\%}~{\rm fb}, \nonumber \\
\kappa_\lambda = 2: \quad 
\sigma_{tot} & = & 13.2(1)^{+1\%}_{-23\%}~{\rm fb}, \nonumber \\
\kappa_\lambda = 2.4: \quad 
\sigma_{tot} & = & 12.7(1)^{+4\%}_{-22\%}~{\rm fb}, \nonumber \\
\kappa_\lambda = 3: \quad 
\sigma_{tot} & = & 17.6(1)^{+9\%}_{-15\%}~{\rm fb}, \nonumber \\
\kappa_\lambda = 5: \quad 
\sigma_{tot} & = & 83.2(3)^{+13\%}_{-4\%}~{\rm fb}, \nonumber \\
\kappa_\lambda = 10: \quad 
\sigma_{tot} & = & 579(1)^{+12\%}_{-4\%}~{\rm fb}
\label{eq:siglam_mt}
\end{eqnarray}
The uncertainties originating from the scheme and scale choice of the
top mass turn out to develop a mild dependence on $\kappa_\lambda$ as
expected. The size of the total uncertainty band is much less sensitive
to $\kappa_\lambda$ than the location of the band. Combining these
relative uncertainties with the previous renormalization and
factorization scale uncertainties of Eq.~(\ref{eq:siglam_rf}) {\it
linearly} we arrive at the central values with combined uncertainties,
\begin{eqnarray} 
\kappa_\lambda = -10: \quad 
\sigma_{tot} & = & 1680^{+13\%}_{-14\%}~{\rm fb}, \nonumber \\
\kappa_\lambda = -5: \quad 
\sigma_{tot} & = & 598.9^{+13\%}_{-15\%}~{\rm fb}, \nonumber \\
\kappa_\lambda = -1: \quad 
\sigma_{tot} & = & 131.9^{+11\%}_{-16\%}~{\rm fb}, \nonumber \\
\kappa_\lambda = 0: \quad 
\sigma_{tot} & = & 70.38^{+8\%}_{-18\%}~{\rm fb}, \nonumber \\
\kappa_\lambda = 1: \quad 
\sigma_{tot} & = & 31.05^{+6\%}_{-23\%}~{\rm fb}, \nonumber \\
\kappa_\lambda = 2: \quad 
\sigma_{tot} & = & 13.81^{+3\%}_{-28\%}~{\rm fb}, \nonumber \\
\kappa_\lambda = 2.4: \quad 
\sigma_{tot} & = & 13.10^{+6\%}_{-27\%}~{\rm fb}, \nonumber \\
\kappa_\lambda = 3: \quad 
\sigma_{tot} & = & 18.67^{+12\%}_{-22\%}~{\rm fb}, \nonumber \\
\kappa_\lambda = 5: \quad 
\sigma_{tot} & = & 94.82^{+18\%}_{-13\%}~{\rm fb}, \nonumber \\
\kappa_\lambda = 10: \quad 
\sigma_{tot} & = & 672.2^{+16\%}_{-13\%}~{\rm fb}
\label{eq:siglam}
\end{eqnarray}
These final numbers should serve as the recommended values for the total
cross sections and uncertainties at the LHC with $\sqrt{s}=13$ TeV as a
function of $\kappa_\lambda$.

\section{Conclusions}
We have analyzed the combination of the usual renormalization and
factorization scale uncertainties of Higgs-pair production via gluon
fusion with the uncertainties originating from the scheme and scale
choice of the virtual top mass in the Yukawa coupling and the
propagators. Due to the observation that the latter relative
uncertainties are nearly independent of the renormalization and
factorization scale choices, the proper combination of the relative
uncertainties is provided by a {\it linear} addition. Our
procedure does not estimate the full uncertainties at NNLO but those at
approximate NNLO without the knowledge of the complete NNLO top-mass
effects.

In a second step we derived the dependence of the uncertainties related
to the top-mass scheme and scale choice on a variation of the trilinear
Higgs self-coupling $\lambda$. The relative uncertainties are again
observed to develop only a small dependence on $\lambda$. We combined
all the uncertainties for $\sqrt{s}=13$ TeV with the ones of the present
recommendation of the LHC Higgs Working Group, obtaining state-of-the-art predictions
for Higgs pair production cross sections at the LHC including both
renormalization/factorization scale and top-quark scale and scheme
uncertainties. \\

\noindent
{\bf Acknowledgments} \\
The authors acknowledge discussions with the Di-Higgs subgroup of the
LHC Higgs Working Group. The work of S.G. is supported by the Swiss
National Science Foundation (SNF). The work of S.G. and M.M. is
supported by the DFG Collaborative Research Center TRR257 ``Particle
Physics Phenomenology after the Higgs Discovery''. F.C. and J.R.
acknowledge financial support by the Generalitat Valenciana, Spanish
Government, and ERDF funds from the European Commission (Grants No.
RYC-2014-16061, No. SEJI-2017/2017/019, No. FPA2017-84543-P, No.
FPA2017-84445-P, and No. SEV-2014-0398). We acknowledge support by the
state of Baden-W\"urttemberg through bwHPC and the German Research
Foundation (DFG) through Grant No. INST 39/963-1 FUGG (bwForCluster
NEMO).

\end{document}